\documentclass[aps,prb,reprint,superscriptaddress,showpacs]{revtex4-1}
\usepackage{graphicx,color}
\usepackage{amsmath,amssymb,float}

\begin{document}

%\preprint{}

\title{Localized magnetic plasmons in all-dielectric $\mu<0$ metastructures }%(mesoscale structures)}


\author{R. Paniagua-Dom\'{\i}nguez}
\email[]{Present address:  Data Storage Institute, Agency for Science, Technology and Research, 
117608 Singapore.}
\affiliation{Instituto de Estructura de la Materia,
Consejo Superior de Investigaciones Cient{\'\i}ficas, Serrano 121,
28006 Madrid, Spain}
\author{L. S. Froufe-P\'erez}
\email[]{Present address: Physics Department,
University of Fribourg, 
CH-1700 Fribourg, Switzerland.}
\affiliation{Instituto de Estructura de la Materia,
Consejo Superior de Investigaciones Cient{\'\i}ficas, Serrano 121,
28006 Madrid, Spain}
\author{J. J. S\'aenz}
\affiliation{Condensed Matter Physics Dept. and Centro de Investigaci\'on en F\'isica de la 
Materia Condensada (IFIMAC), Universidad Aut\'onoma de Madrid, Fco. Tom\'as y Valiente 7, 28049-Madrid, Spain}
\author{J. A. S\'anchez-Gil}
\email[]{Corresponding author: j.sanchez@csic.es}
\affiliation{Instituto de Estructura de la Materia,
Consejo Superior de Investigaciones Cient{\'\i}ficas, Serrano 121,
28006 Madrid, Spain}
\date{\today}

\begin{abstract}

Metamaterials are known to exhibit a variety of electromagnetic properties non-existing in nature. We show that an all-dielectric (non-magnetic) system consisting of deep subwavelength, high permittivity resonant spheres possess effective negative magnetic permeability (dielectric permittivity being positive and small). Due to the symmetry of the electromagnetic wave equations in classical electrodynamics, localized ``magnetic" plasmon resonances can be excited in a metasphere made of such metamaterial. This is theoretically demonstrated by the coupled-dipole approximation and numerically for real spheres, in full agreement 
with the exact analytical solution for the scattering process by the same metasphere 
with effective material properties predicted by effective medium theory. 
The emergence of this phenomenon as a function of structural order within the metastructures is also studied. 
Universal conditions enabling effective negative magnetic permeability relate subwavelength sphere permittivity and size with critical filling fraction. Our proposal paves the way towards (all-dielectric) magnetic plasmonics, with a wealth of fascinating applications.

\end{abstract}
%\keywords{}

\maketitle

% If in two-column mode, this environment will change to single-column
% format so that long equations can be displayed. Use
%\begin{widetext}
%\end{widetext}

\section{Introduction}

The scattering of electromagnetic (EM) waves from macroscopic media is a classical problem of 
widespread interest throughout the entire EM spectrum, from radio and microwaves, 
%with  increasing frequency 
through the THz, IR, and visible domains towards the high-energy UV and 
x-ray band \citep{83Book_BH,91Book_Nieto},
%The variety of phenomena where EM wave scattering plays a leading role is overwhelming: 
playing a leading role in phenomena such as
radar, lidar, remote sensing, metamaterials, plasmonics, etc.
Not to mention other classical waves (acoustic, seismic) or formally analogous problems 
(electron transport, neutron scattering, etc.). 
Within classical electrodynamics, the scatterers' response is described within macroscopic 
Maxwell equations in terms of a dielectric permittivity $\varepsilon$ and a magnetic 
permeability $\mu$ \citep{83Book_BH}. %98Book_Jackson,93Book_Born}). 
In general, and particularly in the high frequency range, non-magnetic media have been considered, based on the fact 
that most materials found in nature exhibit no magnetic permeability; except for a few theoretical works playing with 
artificial magnetic permeabilities \citep{68Veselago,pinheiro2000PRL}.
Nonetheless, the advent of the so-called metamaterials has made it possible to achieve a variety 
of EM responses; these are artificial materials structured at scales much shorter 
than the wavelength with exotic (effective) $\varepsilon$ and $\mu$. In this manner, not only 
large values of the magnetic permeability are possible, but also negative-$\mu$ metamaterials.
Needless to say, such negative magnetic response is 
crucial in .i.e. fabricating the so called ``left-handed" media (LHM), a novel kind of materials 
predicted to exhibit peculiar electromagnetic properties in which both the permittivity and 
the permeability are negative \citep{68Veselago,00PRL_Pendry_Lens,00PRL_Smith_NIM}.
  
Thus, it seems quite natural to address  in the classical scattering of EM waves the impact of 
media with relative  $\mu\neq 1$ \citep{KerkerJOSA83, Pendry1999}, either positive or negative;
%Although both cases are interesting enough, it can be argued that 
in this context, $\mu<0$ media behaving as magnetic metals can be expected to present a richer associated phenomenology, 
as conventional metals do when compared to usual dielectrics. 
Among this phenomenology, the excitation of surface plasmons, i.e. collective oscillations of the 
conduction electrons, plays a major role due, mainly, to the fact that they concentrate the EM energy 
within subwavelength regions.

%{\color{blue}
Interestingly, while the excitation of propagating surface plasmon polaritons in $\mu<0$ and 
left-handed toy-model materials  \citep{Ruppin2000_PLA,Maradudin2001_SPIE,pinheiro2000PRL,%
Gollub2005,PaniaguaDominguez2010_MM}, 
(some of them explicitly stating the symmetry relations between these and their analog 
in conventional $\varepsilon<0$ metals \citep{Gollub2005,PaniaguaDominguez2010_MM}), 
very few works deal with the localized version of these collective modes \cite{Garcia-CamaraJOSAA08}
in realistic scenarios. 
In fact, most of the studies related to scattering from bodies with $\mu<0$
consider simultaneously $\varepsilon<0$, thus being left-handed 
\citep{Ruppin2000_SSC,Ruppin2004_JPCM,PhysRevB02_Maradudin,OptCom07_Merchiers}. %}
%Regarding those dealing with purely $\mu<0$ structures, the studies of plane wave scattering from small homogeneous spheres 
%\citep{,Geffrin2012} are of particular interest.
In recent years, the fact that the lowest-order resonance found in the extinction spectrum of (realistic) high-refractive-index 
spheres/rings typically presents a definite dipolar magnetic character, has been exploited to achieve magnetic dipole response 
with single particles made of non-magnetic materials \citep{08PRL_Cummer,PhysRevLett08_Li,Merlin2009,Jelinek2010,%
PRB10_Evlyukhin,11OE_G-E,NJP11,Geffrin2012,12SciRep_MagnLight}.
Other approaches have been proposed  based on metallic structures (which indeed suffer from large losses)
such as  dielectric microspheres decorated with metal nanoparticles  
\cite{Muhlig2011,Sheikholeslami2013}.
%aiming at achieving magnetic dipole response.
In any case, despite the amount of theoretical and experimental work devoted to the subject, to the best of our knowledge, 
none of them has proposed  realistic, purely mesoscale structures based on (all dielectric) $\mu<0$ metamaterials in 
which magnetic plasmon effects could be observed.
%Other approaches to obtain effective magnetic dipole based on core-shell metalo-dielectric structures exhibit similar drawbacks 
%\cite{NJP11,arxiv1}.

In this work, we propose as a purely $\mu<0$ ($\epsilon>0$) system (described in Sec. II) a collection of high refractive index (HRI) 
spheres (Sec. III).  Moreover, we show in Sec. IV that, when arranged as a finite metastructure, in particular a metasphere, 
this system may, in turn, support localized magnetic plasmon resonances (LMPRs).  Section V includes a
discussion on the regime of validity throughout the EM spectrum, yielding simple, universal conditions  which connect
permittivity and size of the HRI spheres  with a critical filling fraction.Incidentally, note that similar magnetic localized plasmon 
resonances terminology has been used in Ref.~\citep{PhysRevX14_GarciaVidal}, but in the totally different context.
While here, it is used to describe the excitation of plasmons in effective magnetic plasmas, it is introduced there in the context
of localized spoof plasmons, wherein magnetic resonances arise in the extinction spectrum of Perfect Electric 
Conductor corrugated disks.

\section{Generic $\mu<0$ media: Expected scattering properties}

It is well known that the scattering of EM waves
from a homogeneous sphere accepts an analytical solution which, in the case of illumination by a $x$-polarized plane wave travelling 
along the positive $z$-axis, gives the following form for the scattering and extinction efficiencies:
\begin{eqnarray}
Q_{sca}=\frac{2}{x^{2}}\Sigma_{m}(2m+1)(|a_{m}|^2+|b_{m}|^2),\\
Q_{ext}=\frac{2}{x^{2}}\Sigma_{m}(2m+1)Re(a_{m}+b_{m}),
\end{eqnarray}
with
\begin{eqnarray}
a_n=\frac{\mu\psi_{n}(x)\psi'_{n}(mx)-m\psi_{n}(mx)\psi'_{n}(x)}{\mu\xi_{n}(x)\psi'_{n}(mx)-m\psi_{n}(mx)\xi'_{n}(x)},\\
b_n=\frac{m\psi_{n}(x)\psi'_{n}(mx)-\mu\psi_{n}(mx)\psi'_{n}(x)}{m\xi_{n}(x)\psi'_{n}(mx)-\mu\psi_{n}(mx)\xi'_{n}(x)}
\end{eqnarray}
where $x=2\pi n_{0} R/\lambda$ is the size parameter ($R=D/2$ being the sphere radius and 
$\lambda=2\pi c/\omega$ the  wavelength), 
and $ a_{m},b_{m}$ the Mie (electric, magnetic) multipolar coefficients \cite{83Book_BH}, which are functions of
the refractive index $n=(\varepsilon\mu/\varepsilon_0\mu_0)^{1/2}$  (relative to the surrounding medium).

\begin{figure}[b]
\includegraphics[width=\columnwidth]{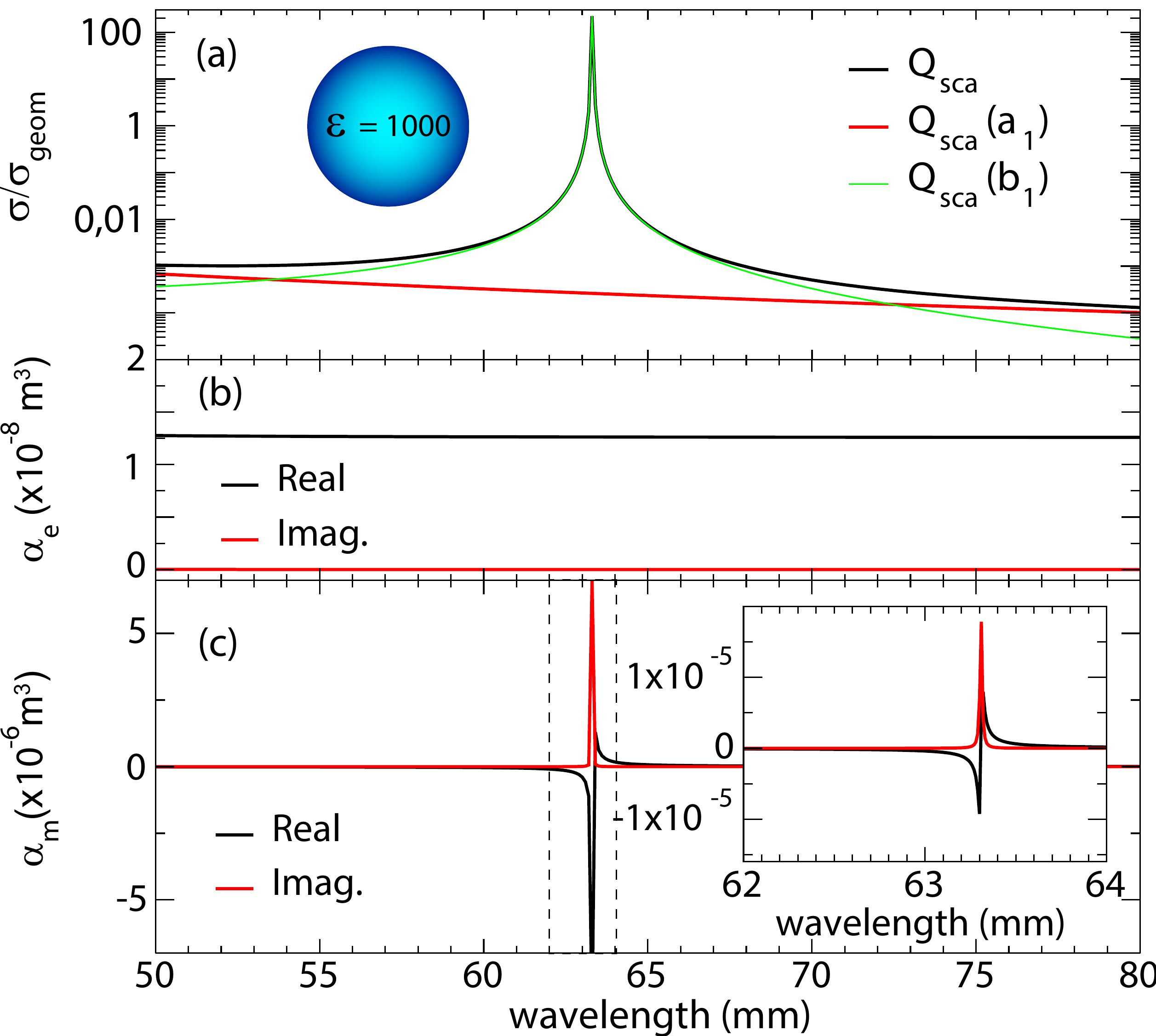}
\caption{Optical properties under plane wave illumination of a sphere with $R=1$~mm, $\varepsilon=1000$ and $\mu=1$. (a) Scattering efficiency (black),
together with the dipolar electric (red) and dipolar magnetic (green) contributions. (b) Real (black) and imaginary (red) parts of the electric
polarizability. (c) Real (black) and imaginary (red) parts of the magnetic polarizability. The inset shows a detail for frequencies around the resonance.}
\label{fig1}
\end{figure} From %the equations that fully describe the EM scattering process, i.e., 
Maxwell's equations 
%{\color{blue}
%and the continuity conditions:
%\begin{subequations}\label{ecs_cc}
%\begin{eqnarray}
%\mathbf{n} \cdot \mathbf{H}^>=\mathbf{n} \cdot \mathbf{H}^<,\ 
%\mu_0 \mathbf{n} \times \mathbf{H}^>=\mu_1 \mathbf{n} \times \mathbf{H}^<\\ 
%\mathbf{n} \cdot \mathbf{E}^>=\mathbf{n} \cdot \mathbf{E}^<,\ 
%\varepsilon_0 \mathbf{n} \times \mathbf{E}^>=\varepsilon_1 \mathbf{n} \times \mathbf{E}^<,
%\end{eqnarray}
%\end{subequations}
%where $\mathbf{n}$ is the unit vector normal to the scatterer surface, 
%}%
it is easy to demonstrate that, upon exchanging 
the value of the relative dielectric permittivity 
$\varepsilon$ with 
that of the relative magnetic permeability 
$\mu$, and viceversa, the 
resulting electric (magnetic) fields are identical to the initial magnetic (electric) fields (except for
a sign change preserving the chirality of the EM waves). In particular, due to the rotational symmetry of this problem, one has:
\begin{eqnarray}
E,\mathbf{H} & \Longleftrightarrow & \tilde{E}=-H,\mathbf{\tilde{H}}=\mathbf{E}\\
\varepsilon, \mu & \Longleftrightarrow & \tilde{\varepsilon}=\mu,\tilde{\mu}=\varepsilon\\
a_{m}, b_{m} & \Longleftrightarrow & \tilde{a}_{m}=b_{m}, \tilde{b}_{m}=a_{m}.
\end{eqnarray}

This symmetry implies that the properties of a magnetic media with $\mu\neq 1,\varepsilon=1$ 
can be inferred from those of a non-magnetic material with $\tilde{\mu}=1,\tilde{\varepsilon}=\mu$. Specifically, under the assumptions 
$\lambda\gg R$ and $\lambda\gg R/\sqrt{\varepsilon\mu}$, the scattering process can be accurately described retaining 
only the first two  terms in Lorenz-Mie expansion \citep{83Book_BH,KerkerJOSA83}, $a_{1}$, $b_{1}$ representing, 
respectively, the dipolar electric and magnetic contributions:
%In the limits previously indicated, they acquire the following form \citep{KerkerJOSA83}:
\begin{equation}\label{eq_dipcoeff}
 a_{1}=\frac{2i}{3}\frac{\varepsilon-1}{\varepsilon+2}x^{3}\ \ \ \ \ \ b_{1}=\frac{2i}{3}\frac{\mu-1}{\mu+2}x^{3}.
\end{equation}
Therefore, for small particles having $\varepsilon=1$ and $\mu\sim-2$ it would be possible, in principle, 
to excite dipolar resonances with analogous properties
to those excited in usual metallic particles (for which $\mu=1$ and $\varepsilon\sim-2$).

\section{High-refractive index spheres as $\mu<0$ metamaterial constituents}

\begin{figure}[h]
\includegraphics[width=\columnwidth]{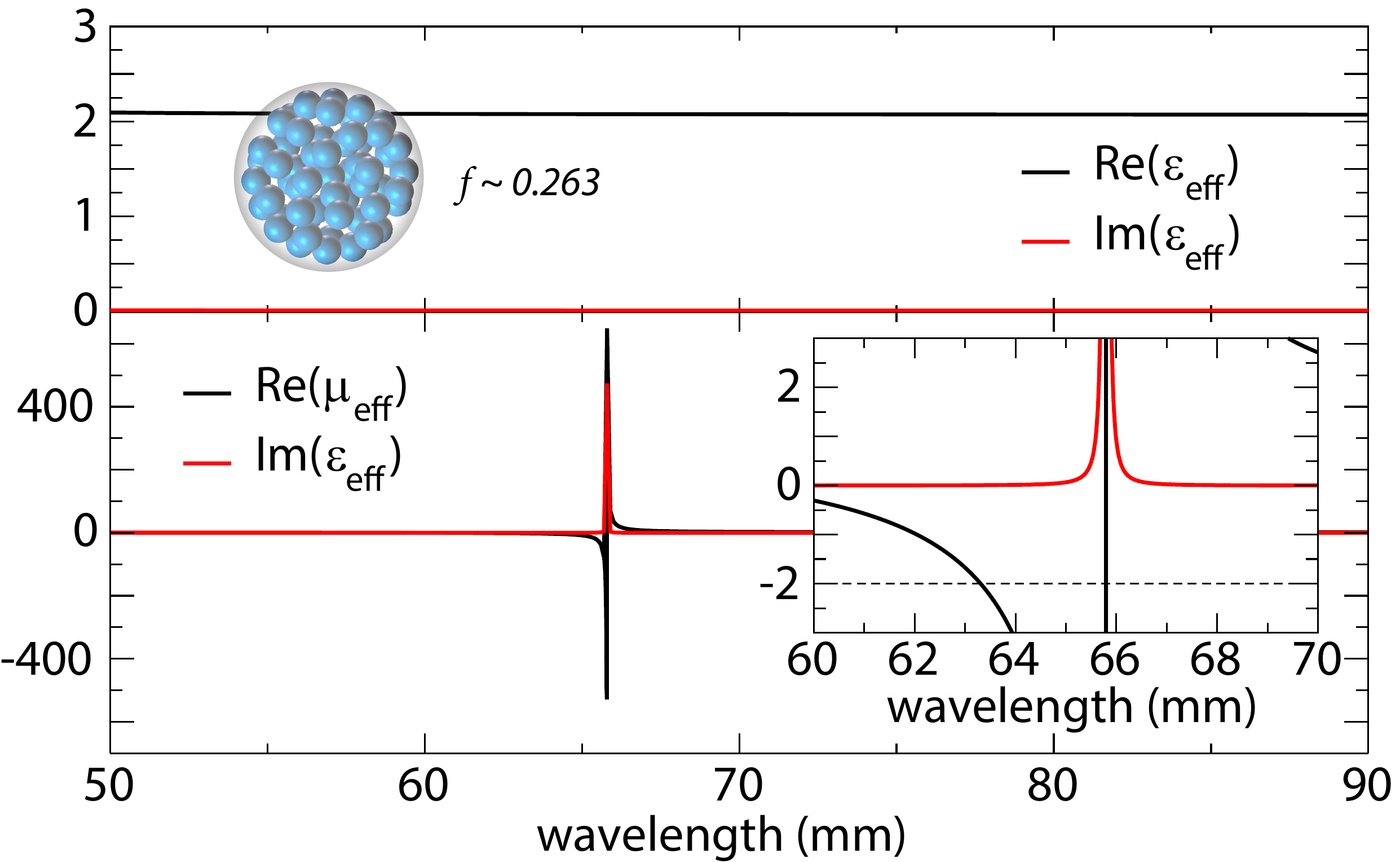}
\caption{Effective material properties of a medium made of the spherical particles of Fig.~\ref{fig1} with a
filling fraction $f=0.263$. (a) Real (black) and imaginary (red) parts of the effective electric permittivity. (b) Real (black) and imaginary (red) 
parts of the effective magnetic permeability.}
\label{fig2}
\end{figure}
The question now arises if there is a medium that fulfills all necessary conditions. 
%In order to address this question
To this end, let us consider 
the optical properties of a HRI non-magnetic sphere ($\varepsilon\gg 1$, $\mu=1$), which presents  
%Particularly well known among the microwave community \citep{08PRL_Cummer,PhysRevLett08_Li,Geffrin2012},
%but quickly extending to people working at optical fequencies \citep{11OE_G-E,12SciRep_MagnLight,PRB10_Evlyukhin}, 
%it is the fact that  in HRI particles the 
a lowest-order resonance 
%found in their extinction spectrum typically presents a definite 
of dipolar magnetic character. 
Moreover, once a certain value of $\varepsilon$ is reached, the position of this resonance scales as 
$\lambda_{0}^{\mathrm{(m)}}\sim 2\sqrt{\varepsilon}R$ ($nx=\pi$),
which opens the possibility to use these HRI systems as building blocks of effective magnetic media. 
For large enough $\varepsilon$ values, $\lambda^{\mathrm{(m)}}\gg R$, which makes the system homogenizable 
and, at the same time, such that $\lambda_1^{\mathrm{(m)}}\gg\lambda_{m}^{\mathrm{(m)}}$, for any other
resonance. That makes the individual building blocks be accurately described as purely magnetic dipoles, 
and the effective medium so obtained to have $\mu_{\mathrm{eff}}\neq1$, but $\varepsilon_{\mathrm{eff}}\sim1$. 
Thus, let us consider a sphere of $R=1$~mm with $\varepsilon=1000$, which is a reasonable value for certain ceramic
materials such as the BST operating in the gigahertz range \citep{PhysRevLett08_Li}:
% and let us fix the radius of a sphere of such material to $R=1$~mm. 
%The optical properties of this system under plane wave illumination are depicted in 
%Fig.\ref{fig1}. In the top panel, 
its scattering efficiency is plotted in Fig.~\ref{fig1}(a), together with the electric and magnetic dipolar 
contributions. A magnetic dipole resonance can be observed at a wavelength 
$\lambda_1^{\mathrm{(m)}}/R\sim2\sqrt\varepsilon\sim 63$, as expected. 
%The multipole decomposition allows us to identify its dipolar magnetic character. 
Moreover, the electric contribution is almost 
negligible, as can be also appreciated in the polarizabilities:
\begin{equation}\label{eq_polariz}
 \alpha_e=6\pi ia_{1}/k^3,\ \ \ \alpha_m=6\pi ib_{1}/k^3,
\end{equation}
shown in Figs.~\ref{fig1}(b-c), the non-resonant electric dipole polarizability being up to two orders of magnitude smaller
than the (resonant) magnetic one. Since the sum of the remaining, higher-order contributions is about five orders of magnitude 
lower, we conclude that each of the HRI spheres can be accurately described retaining only the dipolar contributions, 
namely, by means of a pair of electric and magnetic dipoles.
Incidentally, note that toroidal multipoles have also been introduced elsewhere for arbitrary sources\cite{Radescu2002}, 
and explicitly accounted for in the case of HRI spheres (toroidal dipole) to explain transparency \cite{arxiv1,arxiv2}; 
however, the response of our HRI spheres is fully accounted for (and understood) through their Mie lowest-order magnetic 
($b_1$) contribution, to which, unlike to the Mie electric ($a_1$) one,  the toroidal dipole moment does not contribute 
[as defined in Eq.~(7.24) of Ref.~\cite{Radescu2002} or in Eq.~(3) of Ref.~\cite{arxiv2}, see also Eq.~(\ref{Eq_TD}) below].

Let us now build an effective medium with a collection of these HRI particles. Since the size of the particles 
is extremely subwavelength and the dipolar aproximation holds, we expect the usual effective medium theories 
to give accurate results for the effective properties of the medium, at least, away from the resonance and for low 
volume filling fractions, $f$. Incidentally, it has been recently reported that the effective optical response of colloidal particles, 
in general, may exhibit a non-local magnetic permeability \citep{PhysRevB07_Barrera,JCP14_Barrera}; 
despite that, as a first estimate, we make use of Clausius-Mossotti relations.

The results so obtained for a medium with a filling fraction of $f\sim0.263$ (the election of such number will be clarified afterwards) 
are shown in Fig.~\ref{fig2}. While the effective permittivity of the system remains approximately constant with low values 
($\varepsilon_{\mathrm{eff}}\sim2$),
%the case of 
the magnetic permeability 
%completely differs. It 
shows a resonant behaviour reaching negative values (extremely large in magnitude)
for frequencies above the magnetic resonance of the individual building blocks. Interestingly, although losses around the 
resonance tend to be very high, in the region in which $\mathrm{Re}(\mu)\sim-2$, one finds that both 
$\mathrm{Im}(\varepsilon_{\mathrm{eff}})\sim \mathrm{Im}(\mu_{\mathrm{eff}})\sim 0$. 
Moreover, the scattering cross section (SCS) of the individual constituents at that frequency is only about three times 
the geometrical one ($\pi R^2$). Thus, one could expect that the effective medium theory can still hold, giving accurate results without 
further refinements \citep{PhysRevB07_Barrera,JCP14_Barrera,JOSA06_Guerin}, as we will show below.

\section{$\mu<0$ Metasphere: Magnetic localized plasmon resonance}

\subsubsection{$\mu<0$ effective medium metasphere: Mie scattering}

One may wonder now, what would be the optical properties of a small particle made of such effective medium? 
In order to address this question, let us choose a sphere with these effective properties (those of Fig.~\ref{fig2}) and radius
$R_{{\mathrm{eff}}}$. If $R_{{\mathrm{eff}}}$ is small enough, recalling expressions (\ref{eq_dipcoeff}) for the
dipolar contributions in the small particle limit, we expect a resonance appearing at frequencies $\omega_1^{(\mu)}$ 
such that $\mathrm{Re}(\mu_{\mathrm{eff}}(\omega_1^{(\mu)}))\sim-2$. For this particular system, if we choose 
$R_{{\mathrm{eff}}}=6$~mm, the corresponding wavelength at resonance still fulfills $\lambda_1^{(\mu)}\gg 
\sqrt{\mu_{\mathrm{eff}}\varepsilon_{\mathrm{eff}}}R_{\mathrm{eff}}>R_{\mathrm{eff}}$. Therefore, we
expect an(effective) dipolar magnetic resonance, together with an small electrical contribution going as $Q_{sca}(a_1)\sim x^{4}/6$. 
Additionally, based on what is known for small metallic particles and considering the electric size of the sphere, 
one could also expect the emergence of a quadrupolar magnetic contribution for frequencies $\omega_2^{(\mu)}$
above the dipolar resonance, analogous to the electric quadrupole appearing in the spectra of relatively large 
(still deeply subwavelength) metallic spheres. 
In Fig.~\ref{fig3}(a), the analytical (Mie) scattering and extinction spectra are plotted:
%as a function of the wavelength of incidence. 
A relatively broad resonance can be observed at a wavelength such that 
$\mu_{\mathrm{eff}}=-3.262+0.014i$, together with a narrow peak at smaller wavelengths. Multipolar decomposition 
allows a direct identification of the character of these resonances; as depicted, the broader at $\lambda_1^{(\mu)}=64$~nm 
being the magnetic dipole, the sharper at $\lambda_2^{(\mu)}=63$~nm corresponding to the magnetic quadrupole. 
For larger wavelengths a large number of resonances appear together, with higher order contributions playing 
an important role in the total extinction efficiency of the system. This is due to the fact that the effective permeability 
acquires extremely high positive values once it flips sign, hence making the effective index of refraction also large and 
allowing excitation of such collection of geometrical resonances, some of them, in turn, closely connected to
those present in high permittivity particles.

All the previous would be nothing but a mere theoretical toy-model if there were no realistic system that could exhibit 
this sort of behaviour. Although a practical realization of HRI spheres (of, e.g., BST) is feasible for the range of geometrical 
parameters studied here \citep{PhysRevLett08_Li}, the question of whether or not a collection of them will behave as 
an effective magnetic plasma, as expected from the effective medium theory, is still open. We tackle this problem by a direct 
comparison of the analytical predictions for the extinction of Fig.~\ref{fig3}, calculated using Mie theory with the effective 
parameters of Fig.~\ref{fig2}, with simulations of the full scattering process, involving the whole set of HRI particles. 

\subsubsection{Metasphere made of HRI spheres: Coupled electric/magnetic dipole approach, and full numerical simulations}

\begin{figure}[H]
\includegraphics[width=\columnwidth]{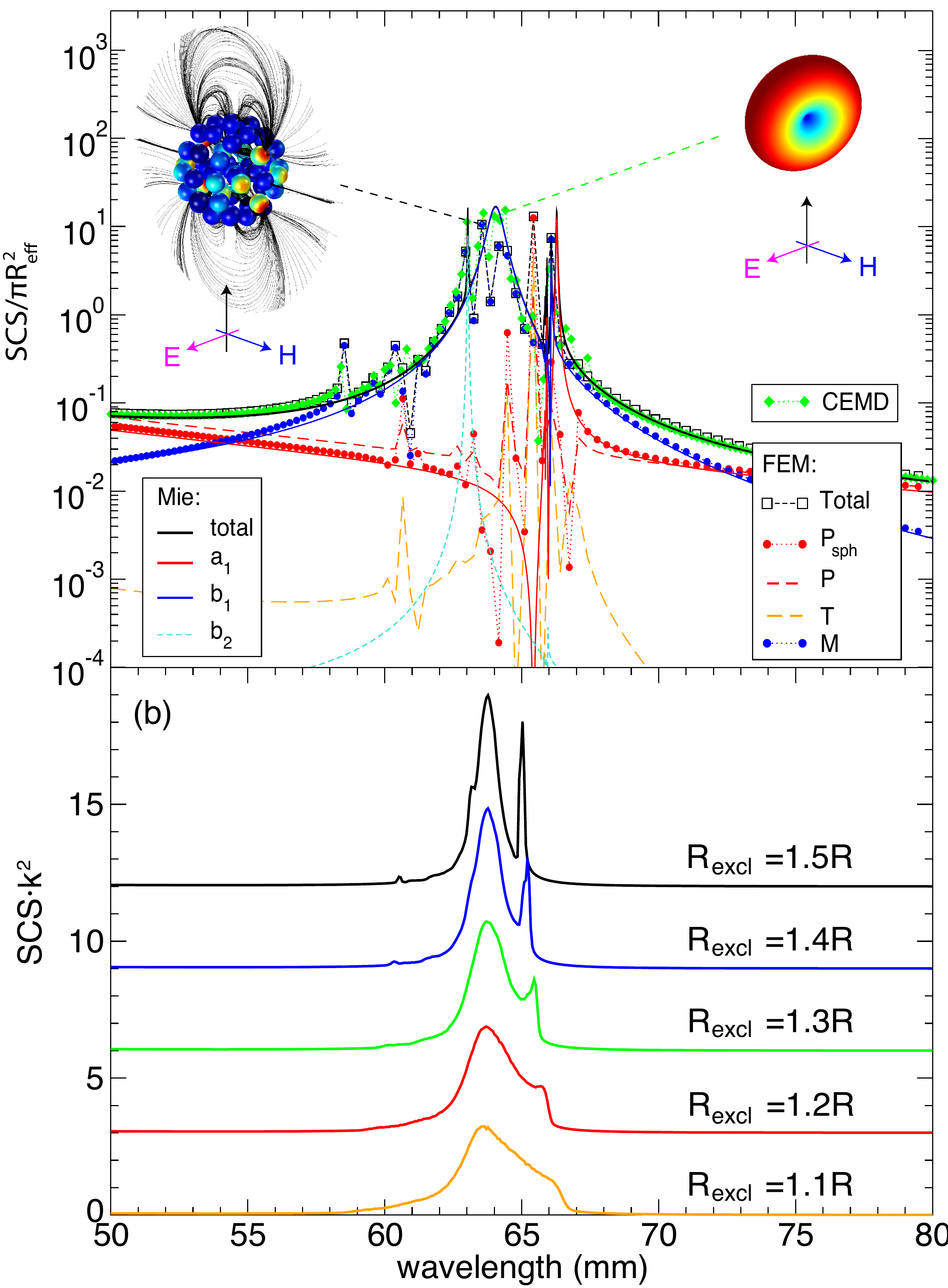}
\caption{(a) Mie theory scattering efficiency (black solid curve) of a sphere with  radius 
$R_{\mathrm{eff}}=6$~mm and material properties $\varepsilon=\varepsilon_{\mathrm{eff}}$ and 
$\mu=\mu_{\mathrm{eff}}$ taken from Fig.~\ref{fig2}, including separately the contributions from the electric ($a_1$, red solid curve) 
and magnetic ($b_1$, blue solid curve) dipole, and the magnetic ($b_2$, cyan dashed curve) quadrupole terms; 
together with the scattering cross section (green dashed curve with diamonds), numerically computed using the CEMD method, 
of $N=57$ distributed electric-and-magnetic dipoles with polarizabilities taken from Fig.~\ref{fig1}(b) and (c). 
Full numerical calculation considering the HRI spheres is shown as a black dashed curve with hollow squares 
(filling fraction $f=0.263$), including separately  (see text) the cartesian electric $\mathbf{P}$ (red dashed curve), 
magnetic $\mathbf{M}$ (blue dotted curve with circles), and toroidal $\mathbf{T}$ 
(orange dashed curve) dipole contributions; along with the spherical electric dipole contribution 
$\mathbf{P}_{\mathrm{sph}}=\mathbf{P}+\imath k\mathbf{P}$ (red dotted curve with circles).
The insets show the CEMD computed scattering pattern (right) 
and the FEM computed magnetic field (lines and amplitude at the surface of the spheres) at the maximum of the 
magnetic resonance (left). (b) CEMD computed SCS for $N_{rea}=10^{5}$ realizations of random arrangements of 
dipoles with several exclusion radii, $R_{\mathrm{excl}}$.}
\label{fig3}
\end{figure}

We do it numerically in two alternative ways. First, through the Coupled Electric and Magnetic Dipoles (CEMD) method 
\citep{Langmuir94CEMD}, which describes each dielectric particle as a pair of coupled electric and magnetic dipoles with 
polarizabilities given by Eqs.~(\ref{eq_polariz}), the magnetic dipole being at resonance. 
The problem is self-consistently solved by expressing the electric and 
magnetic fields at the position of each dipole as a superposition of the incident field and the field generated by the rest of dipoles. 
Once the fields are known at each dipole position the far-field radiated is computed, together with the total SCS. Second, 
through full rigorous numerical calculations based on finite element method (FEM). Both approaches take fully into account 
inter-particle coupling. The results for the CEMD method in a single realization are shown as a blue curve (with circles) 
in Fig.~\ref{fig3}(a). The particular distribution of HRI spheres inside the spherical volume of radius $R_{\mathrm{eff}}$ 
for this realization is shown as an inset in the figure (efficiently occupying the volume of the equivalent sphere), and we assume the 
pair of dipoles to be located in the center of each HRI sphere. In this realization, we consider $N=57$ particles, 
which makes a volume filling fraction $f=0.263$.

Interestingly, even for a single realization, the result of CEMD closely reproduce that predicted for a negative-$\mu$ 
effective medium with the same radius, aside from the emergence of some kinks that depend on the 
particular realization. Investigation of the radiated far-field pattern at the maximum of the resonance 
$\lambda_1^{(\mu)}\sim 64$ mm (shown as an inset) reveals the magnetic dipole character of this collective 
resonance. Features related with the excitation of the magnetic quadrupole, as well as the set of resonances arising 
at wavelengths such that $\mu_{\mathrm{eff}}\gg 1$, can also be observed in the CEMD simulation. Results of full numerical 
simulations carried out using COMSOL v4.3b (a FEM commercial software) are also shown in Fig.~\ref{fig3}(a) as a 
black-dashed  curve (with squares). These are in very good agreement as well, demonstrating the effect in realistic systems that  
could be experimentally  measured. 

To shed even more light onto the physics underlying the above resonance in the context of the electric, magnetic, 
and toroidal multipoles \cite{Radescu2002}, we calculate numerically the electric $\mathbf{P}$, magnetic $\mathbf{M}$, 
and toroidal $\mathbf{T}$ dipole contributions through:
\begin{eqnarray}
&& \mathbf{P}=\frac{\imath}{\omega}\int d^3r \mathbf{J},\\
&&\mathbf{M}=\frac{1}{2c}\int d^3r (\mathbf{r}\times\mathbf{J}),\\
&&\mathbf{T}=\frac{1}{10c}\int d^3r [(\mathbf{r}\cdot\mathbf{J})\mathbf{r}-2r^2 \mathbf{J}]; \label{Eq_TD}
\end{eqnarray}
with the displacement current $\mathbf{J}$ being related to the electric field inside  by:
\begin{eqnarray}
\mathbf{J}= -\imath\omega\varepsilon_0(\varepsilon -1)\mathbf{E}.
\end{eqnarray}
These contributions to the total far-field (SCS) are included also in Fig.~\ref{fig3}(a). Electric and toroidal dipole contributions 
are negligible throughout most of the LMPR lineshape (except for the  $\mu_{\mathrm{eff}}\gg 1$ region mentioned above); 
note also that  the sum of  these two contributions, called  $\mathbf{P}_{\mathrm{sph}}=\mathbf{P}+\imath k\mathbf{T}$ 
(see Ref.\cite{arxiv1}), accurately describes the Mie dipole  contribution ($Q_{\mathrm{sca}}(a_1)$). Only in the minimum of 
the latter $a_1$ scattering channel, there is a noticeable contribution from $\mathbf{T}$, as expected. 

More importantly, it is evident from Fig.~\ref{fig3}(a) that the main (LMPR) resonance is fully accounted for by the magnetic dipole 
$\mathbf{M}$ contribution, in agreement also with the Mie $Q_{\mathrm{sca}}(b_1)$ term.  Furthermore,
the (left) inset in Fig.~\ref{fig3}(a) showing the magnetic field on the surface of the spheres (red corresponding to high intensity and blue to 
low), alongside with the field lines of the scattered magnetic field, corroborates that the whole system collectively resonates with 
the characteristic pattern of a magnetic dipole, providing further support to our $\mu<0$ effective medium approach for the 
metasphere. 

We now study the emergence of this resonance in connection with structural order in random arrangements of particles. To do so 
we apply the CEMD method and consider realization of $N=57$ dipoles placed randomly with only two restrictions: we set a 
maximum distance from the origin to a dipole site equal to $R_{\mathrm{max}}=R_{\mathrm{eff}}-R$, and an exclusion radius 
between dipoles equal to $R_{\mathrm{excl}}$, meaning that the minimum distance between adjacent dipole positions must be 
$2R_{\mathrm{excl}}$. Increasing $R_{\mathrm{excl}}$ we are able to force dipoles to efficiently occupy the volume of the effective 
sphere. Averaged results for $N_{rea}=10^{5}$ realizations and different exclusion radii are shown in Fig.~\ref{fig3}(b). A clear 
signature of the resonance is obtained even in the case of lowest ordering. Nevertheless, the resonance is better resolved (and the 
quadrupole peak starts to pop up) when a certain ordering is imposed. Since variations in the shape of structures supporting 
localized plasmons are known to strongly affect their spectral features, we attribute this effect to the fact that, in the latter case,  
dipoles are located in such a way that the shape of the equivalent sphere is better preserved in the random realizations.

\section{LMPR in $\mu<0$ metamaterials throughout the EM spectrum: regime of validity}

\begin{figure}[h]
\includegraphics[width=\columnwidth]{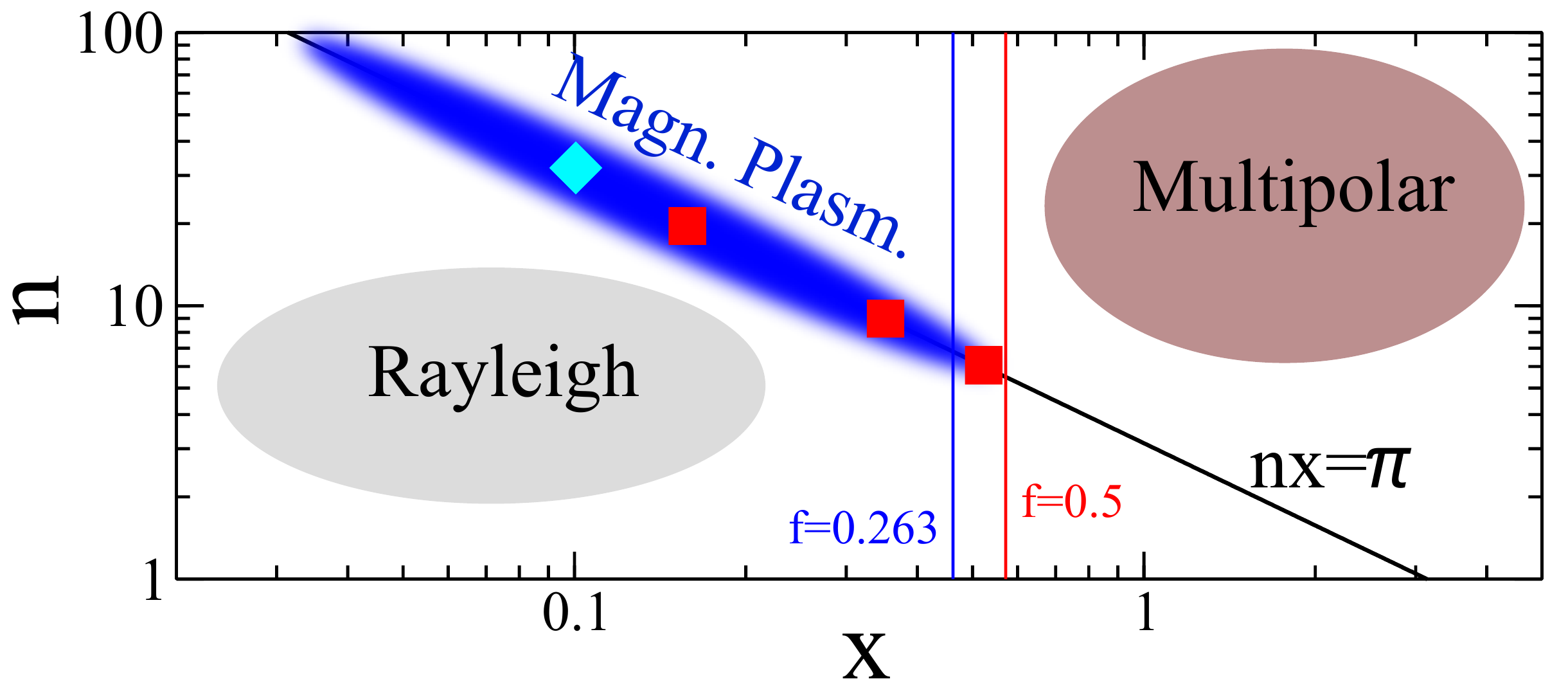}
\caption{Schematic depicting the region of negative permeability ($\mu_{\mathrm{eff}}\ll -1$) in the parameter space $(x,n)$, shown 
as a rotated ellipsoid. The oblique back line identifies the asymptotic (Mie) magnetic-dipole resonance ($nx=\pi$, valid for $n>2.5$),  
predominating over any other resonance, in between the Rayleigh region $x\ll 1, nx\ll 1$ (where the dipolar electric term is 
dominant)  and the  region $x\gg 1, nx\gg 1$, wherein higher-order resonances are larger. Two lower limits of validity are 
established for two filling fractions $f=0.265,0.5$, below which the magnetic permeability becomes $\mu_{\mathrm{eff}}> -1$ at 
half-resonance according to Clausius-Mosotti formula. Squares denote realistic configurations throughout the EM spectrum
(see text), the diamond indicating the particular one studied above.
%, with meta-sphere dimensions extracted from the corresponding $x$: microwave (as considered herein), THz ($n\sim 20$ at 
%$\lambda\sim 200\,\mu$m for SrTiO$_3$), IR ($\lambda\sim 40\,\mu$m for CaF$_2$), and optical ($n\sim 6$ for Ge at 
%$\lambda\sim 0.6\,\mu$m).
}
\label{fig4}
\end{figure}
Finally, let us discuss the emergence of similar $\mu <0$ regimes throughout the entire EM spectrum, inferred from simple 
analytical  constraints imposed on refractive index $n$ and size parameter $x$ by the underlying physics (see Fig.~\ref{fig4}). 
From Mie theory, as mentioned above, it follows that the first magnetic-dipole resonance appears at $nx=\pi$  and 
predominates over any other electric/magnetic resonance (asymptotically for large $n$, though fulfilled from $n> 2.5$, 
as shown in Ref.~\cite{11OE_G-E}); the Rayleigh region $x\ll 1, nx\ll 1$ is shown in the lower left corner, whereas in the region 
$x\gg 1, nx\gg 1$ higher-order resonances prevail.  
Interestingly, a (qualitatively) similar  resonance condition was exploited in Ref.~\citep{Merlin2009} to predict high-frequency 
(positive) magnetic permeability from subwavelength, large-permittivity rings, with the emphasis placed therein on plasmonic 
metamaterials, losses precluding negative effective permeabilities.
This resonance condition is necessary, but not sufficient: the resonance 
strength must be large enough so that $\mu_{\mathrm{eff}}> -1$ at a reasonable filling fraction. We thus impose that 
$\mu_{\mathrm{eff}}=-1$ in the Clausius-Mosotti formula with magnetic polarizability $\alpha_m$ given by 
expression~(\ref{eq_polariz}), with $b_1=\imath$  (cf. Ref.~\cite{11OE_G-E}; in fact,  $\alpha_m/2$ is used instead since the minimum of 
Re$(\mu_{\mathrm{eff}})$ (with Im$(\mu_{\mathrm{eff}})\sim 0$) is expected at a slightly higher energy (half-resonance), 
leading to a condition involving only $f,x$. 

Summarizing both  conditions: 
\begin{eqnarray}
nx=\pi, fx^{-3}=fn^3/\pi^3>4/3.
\end{eqnarray}
The region where $\mu_{\mathrm{eff}}\ll -1$ is depicted in Fig.~\ref{fig4} as an elongated  ellipsoid, without upper limit except for the 
available (natural) refractive index $n$. The particular case investigated above ($n\sim 10\sqrt{10}$ at $\lambda\sim 63$ mm for 
BST in the microwave regime) is marked therein. Moreover, other realistic scenarios where our predicted for metaspheres LMPR  
can be observed in different spectral regimes are also marked  in Fig.~\ref{fig4}: THz ($n\sim 20$ at $\lambda\sim 200\,\mu$m for 
SrTiO$_3$) and IR ($n\sim 9$ at  $\lambda\sim 40\,\mu$m for CaF$_2$). In the visible, Germanium exhibits a refractive 
index $n\sim 6$  at  $\lambda\sim 0.6\,\mu$m, which requires a sphere of radius $R=\lambda/(2n)\sim\lambda/12\sim 50$ nm to 
achieve $\mu_{\mathrm{eff}}\lesssim -1$ with $f\sim 0.4$. Thus the LMPR for a metasphere might be barely achievable in the optical 
domain, but effective medium properties with $\mu_{\mathrm{eff}}<0$ are at reach. 
In this regard, recall that other electric (and thus, magnetic)  plasmonic phenomena do not require effective permittivities 
(permeabilities) smaller than -2 and spheres with $R_{\mathrm{eff}}\ll\lambda$: i.e. higher-order plasmonic resonances either at 
simple geometries for which $\mu_{\mathrm{eff}}\sim-1$ would suffice such as cylinders \cite{83Book_BH}, 
and/or at more complicated structures where geometry plays a major role, like nanorods \cite{NL2014}. 
Thus magnetic plasmonics could also be achieved in the optical domain.
%though alternative approaches based on i.e. HRI meta-atoms could be envisioned.
%In this case, metaspheres of the order of ..., consisting of spheres  of radius $R~$ would yield a LMPR at $\lambda\sim$.  

\section{Conclusions}

In summary, we have theoretically and numerically demonstrated the existence of the so called Localized Magnetic  Plasmon 
Resonances in $\mu<0$ effective media made of all-dielectric (HRI) meta-atoms. % (effective magnetic plasmas). 
We have shown that the excitation of these resonances occurs at frequencies such that $\mu_{\mathrm{eff}}\sim -2$ when the 
overall size of the system is small compared with the incident wavelength, as expected from the analytical Mie 
expressions in conjunction with standard effective medium theories. 
A universal condition for the occurrence of such magnetic plasmonic phenomena is in turn developed, revealing that 
available materials and sound size parameters enable it for a wide spectral regime throughout the microwave, THz, far and 
near IR regimes, and nearly for the optical regime if metastructure geometries other than a metasphere are considered. 
To the best of our knowledge, this is the first time in which a realistic system is proposed that could exhibit this behavior. 
It would be now interesting to explore the rich phenomenology analogous to standard $\varepsilon<0$ plasmonics regarding,
in particular, the zoology of magnetic localized plasmon resonances, including coupled metastructures (dimers, trimers, etc.), 
just to mention some. We, moreover, foresee that the same effective medium could be used to test the excitation of Magnetic 
Surface Plasmon Polaritons on flat surfaces. Our theoretical work thus paves the way towards a realistic $\mu<0$ Plasmonics 
with the wealth of applications of ``electric" plasmonics expected reciprocally in its ``magnetic" counterpart.

\begin{acknowledgments}
We are grateful to  M. Nieto-Vesperinas and F. Moreno for fruitful discussions and critical reading of the manuscript. This work was 
supported by the Spanish MINECO (FIS2012-31070 and FIS2012-36113) and Consolider-Ingenio EMET (CSD2008-00066).
\end{acknowledgments}

%\bibliography{Bibliography}
%

\end{document}